\documentclass[12pt]{article}
\usepackage{epsfig}
\usepackage{color}
\usepackage{lineno}

\begin{document}

\begin{center}

\vspace*{1.0cm}

{\Large \bf{Decay scheme of $^{50}$V}}

\vskip 1.0cm

{\bf F.A.~Danevich$^{a,}$\footnote{Corresponding author. {\it
E-mail address:} danevich@kinr.kiev.ua (F.A.~Danevich).},
M.~Hult$^{b}$, D.V.~Kasperovych$^{a}$, V.R.~Klavdiienko$^{a}$,
G.~Lutter$^{b}$, G.~Marissens$^{b}$, O.G.~Polischuk$^{a}$,
V.I.~Tretyak$^{a}$}

\vskip 0.3cm

$^{a}${\it Institute for Nuclear Research of NASU, 03028 Kyiv,
Ukraine}

$^{b}${\it European Commission, Joint Research Centre, Retieseweg
111, 2440 Geel, Belgium}

\end{center}

%\linenumbers

\vskip 0.5cm

\begin{abstract}

Investigation of the $^{50}$V electron-capture to the $2^+$ 1553.8
keV level of $^{50}$Ti and search for $\beta^-$ decay of $^{50}$V
to the $2^+$ 783.3 keV level of $^{50}$Cr (both those decays are
fourfold forbidden with $\Delta J^{\Delta \pi}=4^+$) have been
performed using a vanadium sample of natural isotopic abundance
with mass of 955 g. The measurements were conducted with the help
of an ultra low-background HPGe-detector system located 225 m
underground in the laboratory HADES (Belgium). The measured value
of the half-life of $^{50}$V for electron capture was
$T^{\mathrm{EC}}_{1/2}=(2.77^{+0.20}_{-0.19})\times 10^{17}$ yr.
The $\beta^-$-decay branch was not detected and the corresponding
lower bound of the half-life was $T^{\beta}_{1/2}\geq8.9\times
10^{18}$ yr at the 90\% confidence level.

\end{abstract}

\vskip 0.4cm


\section{INTRODUCTION}

\begin{figure}[ht]
 \begin{center}
 \mbox{\epsfig{figure=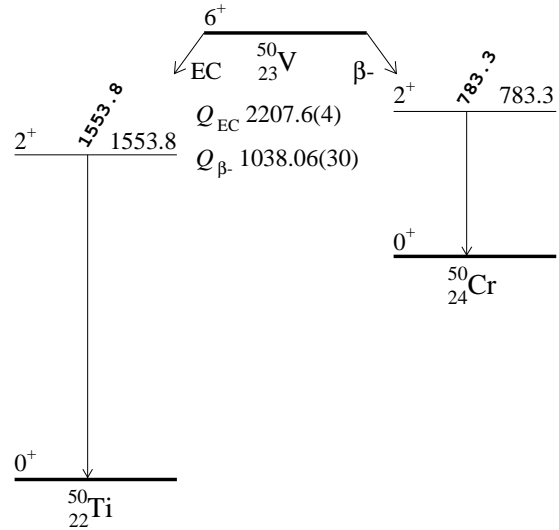,height=7.0cm}}
\caption{Decay scheme of $^{50}$V. No confirmed observation of the
$\beta^-$ decay of $^{50}$V to the $2^+$ 783.3 keV level of
$^{50}$Cr has yet been performed.}
 \label{fig:decay-scheme}
 \end{center}
 \end{figure}

The isotope $^{50}$V is present in the natural mixture of vanadium
with a very low abundance of 0.250(10)\% \cite{Meija:2016}. Taking
into account the mass difference between $^{50}$V and $^{50}$Ti
($2207.6\pm0.4$ keV \cite{Wang:2017}), and between $^{50}$V and
$^{50}$Cr ($1038.06\pm0.30$ keV \cite{Wang:2017}), both electron
capture (EC) of $^{50}$V to $^{50}$Ti and $\beta^-$ decay of
$^{50}$V to $^{50}$Cr are possible (the decay scheme of $^{50}$V
is shown in Fig. \ref{fig:decay-scheme}). However, decays of
$^{50}$V to the ground states of $^{50}$Ti and $^{50}$Cr are
strongly suppressed by the very large spin change $\Delta J = 6$
in both the cases. The only excited levels on which decay of
$^{50}$V can undergo are the $2^+$ 1553.8 keV level of $^{50}$Ti,
and the $2^+$ 783.3 keV level of $^{50}$Cr. Both the decay
channels are fourfold forbidden non-unique ($\Delta
J^{\Delta\pi}=4^+$). Since in both channels decay goes to the
excited levels of daughter nuclei, de-excitation $\gamma$-ray
quanta can be detected by $\gamma$ spectrometry of a vanadium
sample. While the $^{50}$V electron-capture transition to the
$2^+$ 1553.8 keV level of $^{50}$Ti is observed in several
experiments, the $\beta^-$ decay of $^{50}$V to the $2^+$ 783.3
keV level of $^{50}$Cr remains unobserved (despite two claims of
detection that have been disproved in the subsequent more
sensitive investigations). The history of $^{50}$V decays
investigations is summarized in Table \ref{tab:hist} (see also
recent review \cite{Belli:2019}).

The decay of $^{50}$V is of especial interest since the
transitions involve several different nuclear matrix elements with
the associated different phase-space factors multiplied by the
axial-vector coupling constant $g_A$ \cite{Haaranen:2014}. This
constant plays an important role in the neutrinoless double
$\beta$ decay probability calculations
\cite{Barea:2014,DellOro:2014,Suhonen:2017a,Suhonen:2017b,Ejiri:2019}.
Recent calculations in nuclear shell model \cite{Haaranen:2014}
result in the following (partial) half-lives for the two decay
modes: $T^{\mathrm{EC}}_{1/2} = (5.13 \pm 0.07)[(3.63 \pm 0.05)]
\times 10^{17}$ yr given for $g_{A} = 1.00[1.25]$; for the
$\beta^-$-decay branch, $T^{\beta}_{1/2} = (2.34 \pm 0.02)[(2.00
\pm 0.02)] \times 10^{19}$ yr.

\nopagebreak

\clearpage
\begin{table}[htbp]
\caption{Half-lives of $^{50}$V relative to the electron capture
to the $2^+$ 1553.8 keV excited level of $^{50}$Ti and $\beta^-$
decay to the $2^+$ 783.3 keV excited level of $^{50}$Cr.}
\begin{center}
\begin{tabular}{|l|l|l|l|}
 \hline
 Reference (year)               & Experimental technique                & \multicolumn{2}{c|}{Half-life (yr)}\\
\cline{3-4}
 ~                              & ~                                     & Electron capture  & $\beta^-$ decay \\
 ~                              & ~                                     & to $2^+$ 1553.8 keV & to $2^+$ 783.3 keV \\
 \hline
 \cite{Ileintze:1955} (1955)    & Geiger counter,                       & ~                             &   \\
 \cite{Ileintze:1955} (1955)    & Proportional counter                  & $>3.0\times10^{15}$           &  $>3.0\times10^{14}$     \\
 \hline
 \cite{Glover:1957} (1957)      & Proportional counter,                 &  &  \\
                                & NaI(Tl) scintillation counter         & $(4.0\pm1.1)\times10^{14}$    &  $>2.4\times10^{14}$     \\
 \hline
 \cite{Bauminger:1958} (1958)         & Proportional counter,                 &  &  \\
                                & NaI(Tl) scintillation counter         & $(4.8\pm1.2)\times10^{14}$    &  --  \\
 \hline
 \cite{McNair:1961} (1961)      & NaI(Tl) scintillation counter         & $>8.0\times10^{15}$  & $>1.2\times10^{16}$  \\
 \hline

 \cite{Watt:1962} (1962)        & NaI(Tl) scintillation counter         & $(8.9\pm1.6)\times10^{15}$  & $(1.8\pm0.6)\times10^{16}$  \\
 \hline
 \cite{Sonntag:1966} (1966)     & NaI(Tl) scintillation counter         & $>9.0\times10^{16}$  & $>6.9 \times10^{16}$ \\
 \hline
 \cite{Pape:1977} (1977)        & Ge(Li) $\gamma$ spectrometry          & $>8.8\times10^{17}$ & $>7.0\times10^{17}$ \\
 \hline
 \cite{Alburger:1984} (1984)    & HPGe $\gamma$ spectrometry             & $(1.5^{+0.3}_{-0.7})\times10^{17}$ & $>4.3\times10^{17}$ \\
 \hline
 \cite{Simpson:1985} (1985)     & HPGe $\gamma$ spectrometry & $(1.2^{+0.8}_{-0.4})\times10^{17}$ & $>1.2\times10^{17}$  \\
 \hline
 \cite{Simpson:1989} (1989)     & HPGe $\gamma$ spectrometry            & ($2.05\pm0.49)\times10^{17}$ & $(8.2^{+13.1}_{-3.1})\times10^{17}$ \\
 \hline
 \cite{Dombrowski:2011} (2011)  & HPGe $\gamma$ spectrometry            & ($2.29\pm0.25)\times10^{17}$ & $>1.5\times10^{18}$ \\
 \hline
 \cite{Laubenstein:2019} (2019) & HPGe $\gamma$ spectrometry            &  $(2.67^{+0.16}_{-0.18})\times10^{17}$  & $>1.9\times10^{19}$\\
 \hline
 This work   (2020)             & HPGe $\gamma$ spectrometry            & $(2.77^{+0.20}_{-0.19})\times10^{17}$ & $>8.9\times10^{18}$ \\
 \hline
\end{tabular}
\normalsize
\end{center}
\label{tab:hist}
\end{table}

In this work we report measurement of the $^{50}$V EC decay
half-life and search for $\beta^-$ decay of the nuclide using HPGe
$\gamma$ spectrometry of a 955 g vanadium sample.

\section{EXPERIMENT}
\label{sec:exp}

\begin{figure}[ht]
\begin{center}
 \mbox{\epsfig{figure=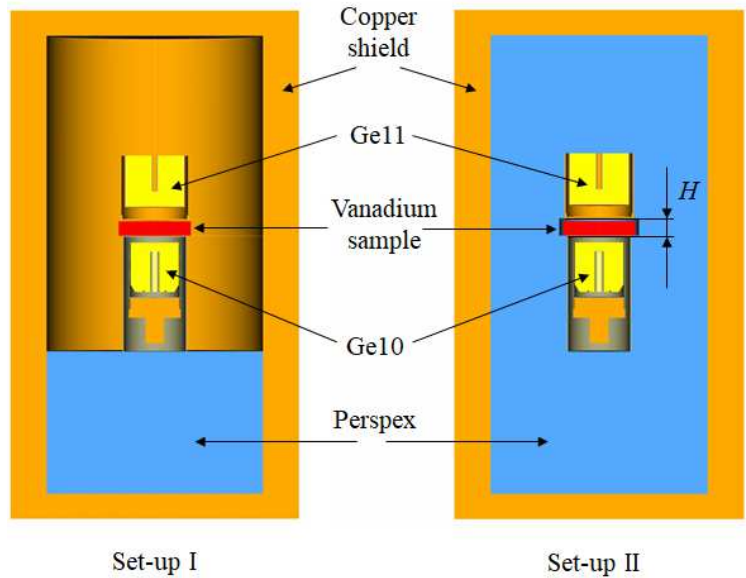,height=8.0cm}}
\caption{(Color online) Schematic view of the inner shield (Pb not
shown) of the two low-background setups with HPGe detectors and
vanadium sample. $H$ denotes distance between the detectors Ge10
and Ge11, that can be adjusted taking into account a sample
height.}
 \label{fig:set-ups}
\end{center}
\end{figure}

A disk-shaped sample of metallic vanadium with diameter of 100.1
mm and thickness of 19.9 mm with mass of $955.21\pm0.02$ g,
provided by Goodfellow Cambridge Ltd was used in the experiment.
The vanadium disk was stored underground as soon as it was
received by JRC-Geel in 2008 so that cosmogenic activation would
be minimized. It was measured using an ultra low-background
HPGe-detector system located 225 m underground in the laboratory
HADES (Belgium). The detector system, named Pacman, consists of
two HPGe-detectors facing each other \cite{Lutter:2013}. The
experiment was realized in two stages with different amount of
Perspex in the inner volume of the lead/copper shield. At the
start not all Perspex was available but due to time constraints it
was judged beneficial to start the measurements anyhow. A
schematic view of the two setups with HPGe detectors and the
vanadium sample is shown in Fig. \ref{fig:set-ups}. The main
characteristics of the HPGe detectors are presented in Table
\ref{tab:detectors}, more details can be found in
\cite{Lutter:2013,Hult:2013}.

\begin{table}[htb]
\caption{Properties of the HPGe-detectors used in the present
experiment. FWHM denotes the full width at half of maximum of
$\gamma$-ray peak. HPAl = High Purity Aluminum. LB Cu = Low
Background Copper}
\begin{center}
\begin{tabular}{|l|c|c|}
 \hline
 ~                                      & Ge10              & Ge11 \\
 \hline

 Energy resolution (FWHM) at 1332 keV   & 1.7 keV           & 1.9 keV \\

 Relative efficiency                    & 62\%              & 85\% \\

 Crystal mass                           & 1040 g            & 1880 g \\

 Endcap / Window material               & HPAl / HPAl       & LB Cu / LB Cu \\

 Other characteristic                   & Submicron outer   & Inverted endcap \\
 ~                                      & deadlayer         & (i.e. the window facing down) \\
 \hline

\end{tabular}
\label{tab:detectors}
\end{center}
\end{table}

\begin{figure}[ht]
\begin{center}
 \mbox{\epsfig{figure=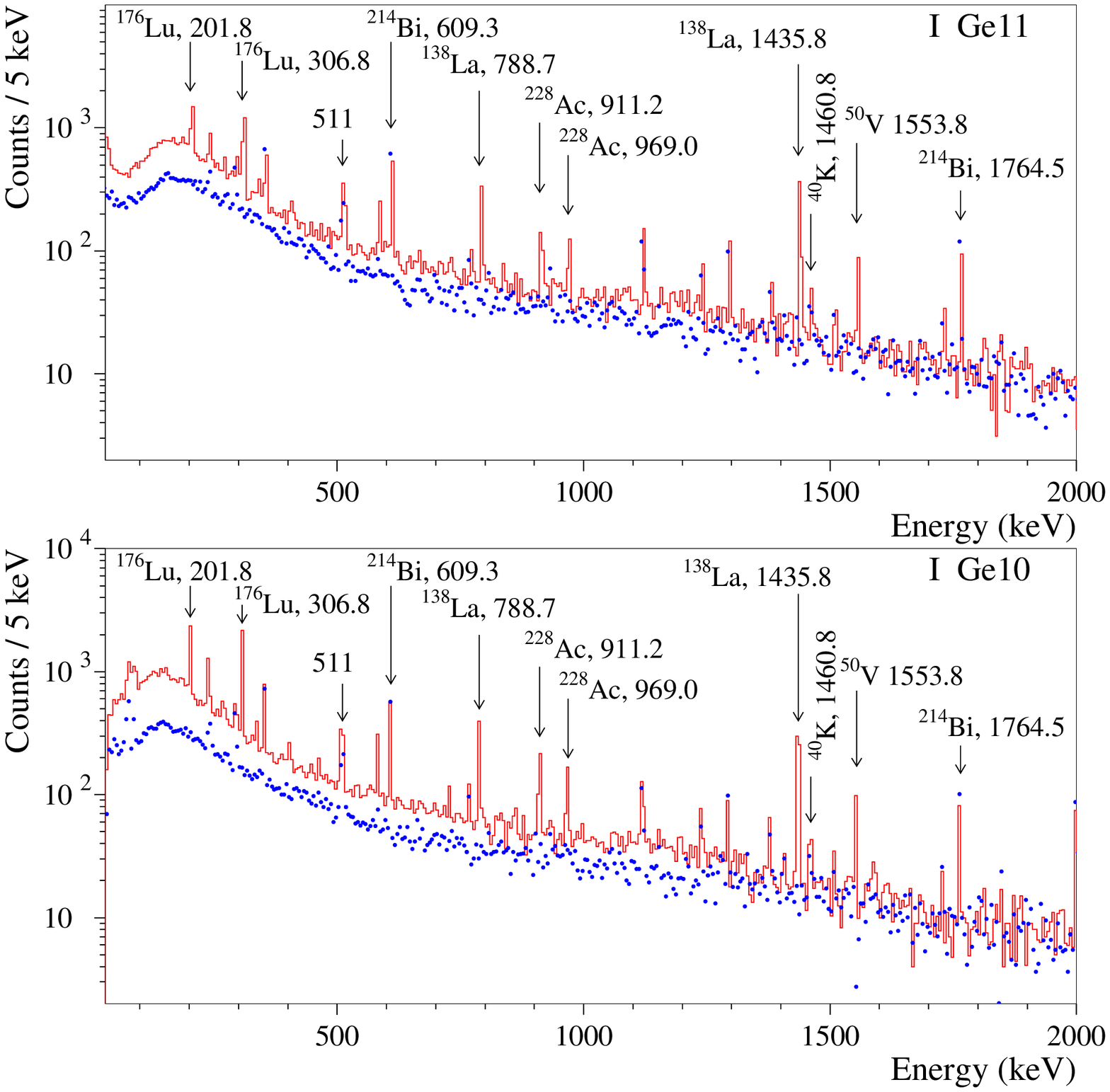,height=13.0cm}}
\caption{(Color online) The energy spectra accumulated in setup I
with the vanadium sample for 34.74 days by detectors Ge11 (upper
panel) and Ge10 (lower panel) (solid lines). The dotted histograms
show background data measured without sample for 38.16 days by the
detector Ge11 (upper panel) and Ge10 (lower panel). The background
spectra are normalized on the time of measurements with the
sample. Energy of $\gamma$-ray peaks are in keV.}
 \label{fig:set-up-I}
\end{center}
\end{figure}

At the first stage of the experiment in setup I the vanadium
sample was measured for 34.74 d, then the detectors were running
for 38.16 d to measure background data without sample. The
distance between the detectors Ge10 and Ge11 was 21 mm in setup I.
The energy spectra accumulated with the vanadium sample and
without sample in setup I are shown in Fig. \ref{fig:set-up-I}.

\begin{figure}[ht]
\begin{center}
 \mbox{\epsfig{figure=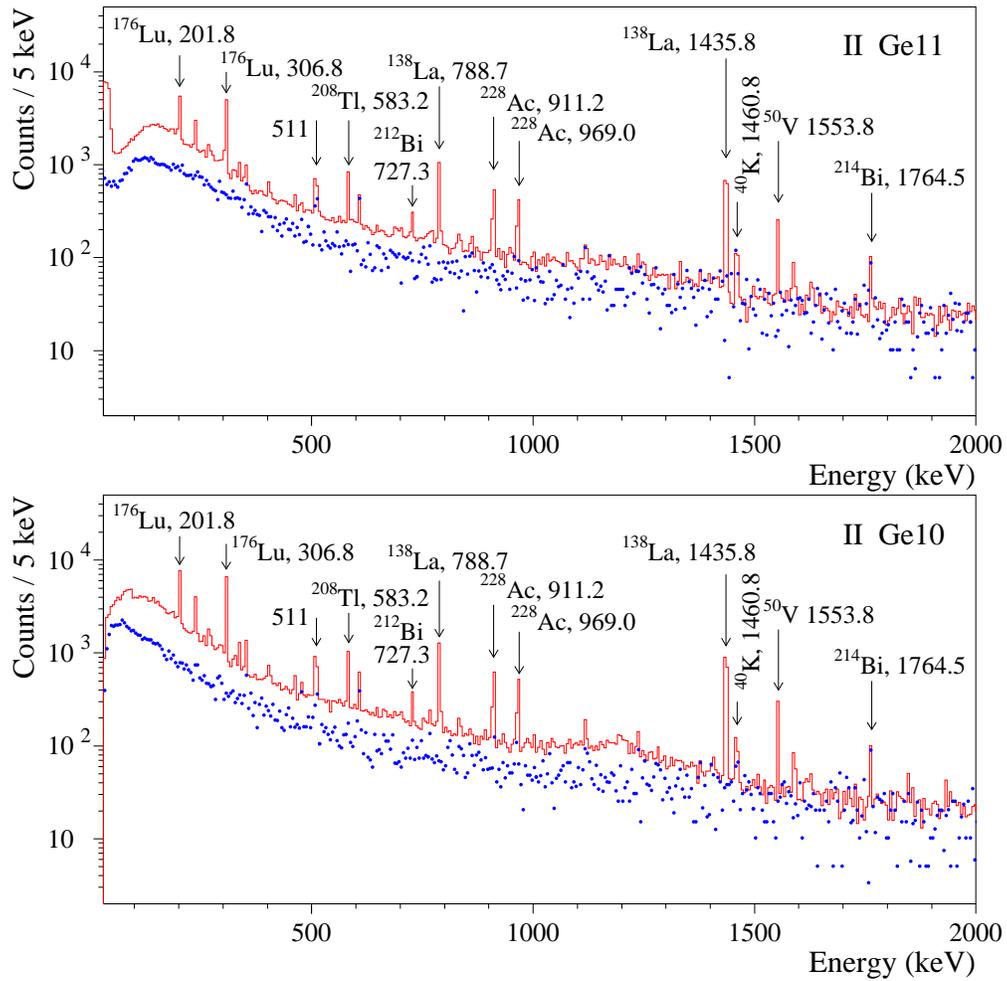,height=13.0cm}}
\caption{(Color online) The energy spectra accumulated in setup II
with the vanadium sample by detectors Ge11 (for 110.55 d, upper
panel) and Ge10 (110.55 d, lower panel) (solid lines). The dotted
histograms show background data measured without sample by
detector Ge11 (21.70 d, upper panel) and Ge10 (21.70 d, lower
panel). The background spectra are normalized on the time of
measurements with the sample. Energy of $\gamma$-ray peaks are in
keV.}
 \label{fig:set-up-II}
\end{center}
\end{figure}

Then the experiment was continued in setup II for 110.55 d with
the vanadium sample and over 21.70 d to measure background without
sample. The distance between the detectors Ge10 and Ge11 was 23 mm
in setup II. Additional Perspex pieces were installed in setup II
to minimize air inside so that to suppress background due to
radon. The energy spectra gathered in setup II are shown in Fig.
\ref{fig:set-up-II}. The insertion of the Perspex details
decreased background caused by $^{222}$Rn daughters. In particular
the counting rates in the $\gamma$-ray peaks of $^{214}$Bi with
energies 609.3 keV and 1764.5 keV were decreased by 3-5 times.

\clearpage

The energy spectra measured in the two setups are rather similar.
The majority of the peaks could be assigned to $^{40}$K and
nuclides of the $^{232}$Th, $^{235}$U, and $^{238}$U decay chains.
Besides, there are also clear peaks of $^{138}$La and $^{176}$Lu
in the data taken with the vanadium sample that is evidence of the
V-sample contamination by La and Lu. No unidentified peaks were
observed.

The energy dependence of the energy resolution in the sum energy
spectrum of the detectors Ge11 and Ge10 in setups I and II was
estimated by using clear $\gamma$-ray peaks with energies
$E_{\gamma}=201.8$ keV and 306.8 keV ($^{176}$Lu), 583.2 keV
($^{208}$Tl), 609.3 keV and 1120.3 keV ($^{214}$Bi), 788.7 keV
($^{138}$La), 911.2 keV ($^{228}$Ac) as ($E_\gamma$ is in keV):

\begin{equation}
 \mathrm{FWHM(keV)}=0.72(9)+\sqrt{0.0019(8)\times E_{\gamma}}-0.0003(2)\times E_{\gamma}.
 \label{eq:fwhm}
\end{equation}

\section{RESULTS AND DISCUSSION}
\label{sec:res-disc}

\subsection{Radioactive impurities in the vanadium sample}
\label{sec:rad-cont}

Massic activities in the vanadium sample of $^{40}$K, $^{138}$La,
$^{176}$Lu, daughters of the $^{232}$Th, $^{235}$U, and $^{238}$U
decay chains were calculated with the following formula:

\begin{equation}
A = (S_{sample}/t_{sample}-S_{bg}/t_{bg})/(\eta~\varepsilon~m),
 \end{equation}

\noindent where $S_{sample}$ ($S_{bg}$) is the area of a peak in
the sample (background) spectrum; $t_{sample}$ ($t_{bg}$) is the
time of the sample (background) measurement; $\eta$ is the
$\gamma$-ray emission intensity of the corresponding transition;
$\varepsilon$ is the full energy peak efficiency; $m$ is the
sample mass. The detection efficiencies were calculated with
EGSnrc simulation package \cite{Kawrakow:2017,Lutter:2018}, the
events were generated homogeneously in the V sample. The
calculations were validated using a liquid solution containing
$^{133}$Ba, $^{134}$Cs, $^{137}$Cs, $^{60}$Co, and $^{152}$Eu. The
standard deviation of the relative difference between the
simulations and the experimental data is $2.5\%$ for $\gamma$-ray
peaks in the energy interval 53 keV--1408 keV for Ge10 detector,
and is 4\% for $\gamma$-ray peaks in the energy interval
80~keV--1408 keV for Ge11 detector. The estimated massic
activities of radioactive impurities in the vanadium sample are
presented in Table \ref{tab:rad-cnt}.

\nopagebreak
\begin{table}[ht]
\caption{Radioactive contamination of the V sample measured by
HPGe $\gamma$-ray spectrometry. The upper limits are given at 90\%
confidence level (C.L.), the reported uncertainties are the
combined standard uncertainties.}
\begin{center}
\begin{tabular}{|l|l|l|}

 \hline
  Chain     & Nuclide      &  Massic activity (mBq/kg) \\
 \hline
 ~          & $^{40}$K     & $3.7\pm1.2$ \\
 ~          & $^{50}$V     & $2.34\pm0.10$ \\
 ~          & $^{138}$La   & $18.7\pm0.2$ \\
 ~          & $^{176}$Lu   & $22.9\pm0.2$ \\
 \hline
 $^{232}$Th & $^{228}$Ra   & $16.1\pm0.6$ \\
 ~          & $^{228}$Th   & $12.7\pm1.0$ \\
 \hline
 $^{235}$U  & $^{235}$U    & $\leq4.9$  \\
 ~          & $^{231}$Pa   & $\leq7.3$  \\
 ~          & $^{227}$Ac   & $11.4\pm0.5$ \\
 \hline
 $^{238}$U  & $^{234m}$Pa  & $41\pm9$ \\
  ~         & $^{226}$Ra   & $\leq0.5$   \\
  \hline
\end{tabular}
\label{tab:rad-cnt}
\end{center}
\end{table}

\clearpage

\subsection{Electron capture decay of $^{50}$V to the $2^+$ 1553.8 keV excited level of $^{50}$Ti}
\label{sec:EC}

\begin{figure}[ht]
\begin{center}
 \mbox{\epsfig{figure=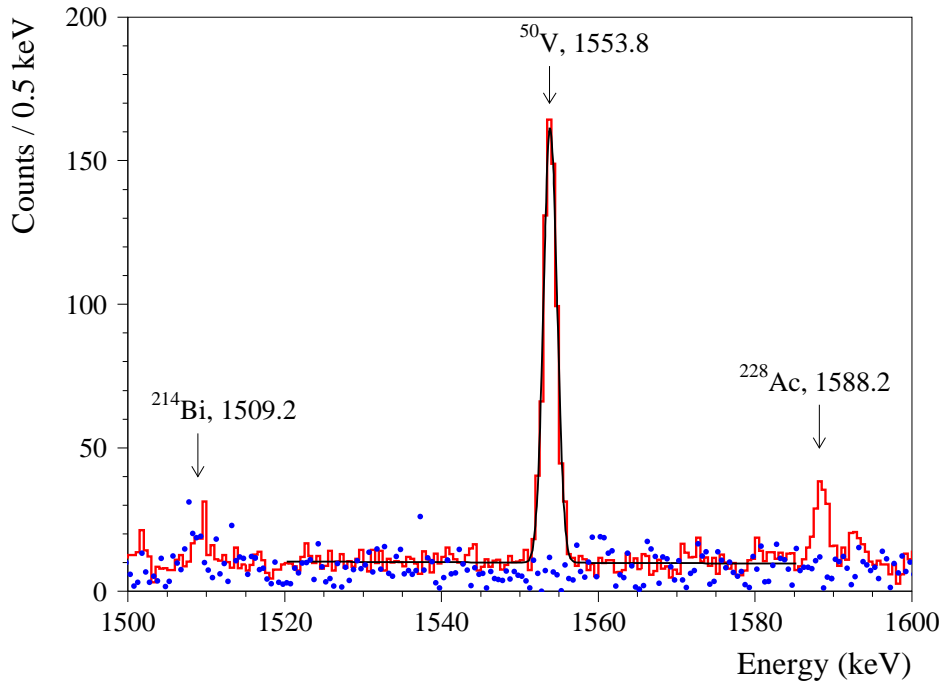,height=9.0cm}}
\caption{(Color online) The sum energy spectrum accumulated with
the V sample in the vicinity of the 1553.8 keV $\gamma$-ray peak
of $^{50}$V. The fit of the data by a sum of Gaussian peak
(effect) and a straight line (background) is shown. The background
energy spectrum, normalized on the time of measurements with the
sample is shown by dots. Energy of $\gamma$-ray peaks are in keV.}
 \label{fig:EC}
\end{center}
\end{figure}

There is a clear peak with energy 1553.8 keV in all the energy
spectra accumulated with the vanadium sample that can be ascribed
to the electron capture decay of $^{50}$V to the $2^+$ 1553.8~keV
level of $^{50}$Ti. The peak is absent in the background data. In
order to estimate the half-life of $^{50}$V for the EC decay
channel the sum energy spectrum of all the measurements with the
vanadium sample was analyzed. A part of the spectrum in the energy
region of interest is presented in Fig. \ref{fig:EC}. The exposure
for $^{50}$V is $(2.25\pm0.09)\times10^{22}$ nuclei of
$^{50}$V$\times$yr.

The spectrum was fitted in the energy interval (1520--1585) keV by
a sum of a first order polynomial function (to describe the
continuous distribution near the peak) and by Gaussian function
(to describe the $\gamma$-ray peak). The fit with a very good
value of $\chi^2/$n.d.f.~=~100.2/126~=~0.795 (where n.d.f. is
number of degrees of freedom) returns the following peak
parameters: energy of the peak is 1553.90(12)~keV (in a good
agreement with the table value 1553.768(8) keV \cite{Chen:2019}),
the FWHM~$=2.02(8)$ keV (again in a good agreement with the
expected FWHM~$=1.95$ keV, see formula (\ref{eq:fwhm})), the area
of the peak is 654(27) counts.

The detection efficiencies for different detectors in the two
setups for $\gamma$-ray quanta with energy 1553.8 keV were
simulated with the help of the EGSnrc package
\cite{Kawrakow:2017,Lutter:2018}. The detection efficiencies are
given in Table \ref{tab:T12-diff}.

\begin{table}[ht]
\caption{Monte Carlo simulated full energy peak detection
efficiencies for 1553.8 keV $\gamma$-ray quanta, live-times of the
measurements, areas of the 1553.8 keV peak, $^{50}$V half-life
values ($T^{\mathrm{EC}}_{1/2}$) for different detectors in the
two setups. The standard statistical errors of the detection
efficiencies, areas of the peak and half-life values are given.}
\begin{center}
\begin{tabular}{|l|l|l|l|l|l|}

 \hline
 Setup & Detector  & Detection efficiency  & Live-time of      & 1553.8 keV    & $T^{\mathrm{EC}}_{1/2}$ \\
 ~      & ~         &                       & measurement (s)   & peak area     & $\times10^{17}$ (yr)\\
 \hline
 I      & Ge11      & $0.011324(25)$        & 3000651           & $79(9)$      & $2.67^{+0.34}_{-0.27}$ \\
 I      & Ge10      & $0.012508(25)$        & 3002755           & $83(10)$     & $2.81^{+0.38}_{-0.30}$ \\
 II     & Ge11      & $0.010607(22)$        & 9551494           & $220(16)$    & $2.86^{+0.22}_{-0.19}$ \\
 II     & Ge10      & $0.012536(25)$        & 9551342           & $270(17)$    & $2.75^{+0.18}_{-0.16}$ \\
 \hline
\end{tabular}
\label{tab:T12-diff}
\end{center}
\end{table}

The half-life of $^{50}$V relative to the electron capture to the
$2^+$ 1553.8 keV level of $^{50}$Ti ($T_{1/2}$) was calculated by
using the following formula:

\begin{equation}
T_{1/2}=N~\ln 2~\sum{(\eta_i~t_i)}/S
 \label{eq:t12}
\end{equation}

\noindent where $N$ is number of $^{50}$V nuclei in the sample
[$N=2.823(113)\times10^{22}$], $\eta_i$ and $t_i$ are detection
efficiencies and times of measurement for the two detectors in the
two setups (given in Table~\ref{tab:T12-diff}), $S$ is area of the
peak with energy 1553.8 keV obtained by the fit of the data of the
sum energy spectrum shown in Fig. \ref{fig:EC} ($S=654\pm27$
counts). By using these data the half-life of $^{50}$V has been
calculated as
$T^{\mathrm{EC}}_{1/2}=[2.774^{+0.119}_{-0.110}(\mathrm{stat})]\times
10^{17}$ yr.

\begin{figure}[ht]
\begin{center}
 \mbox{\epsfig{figure=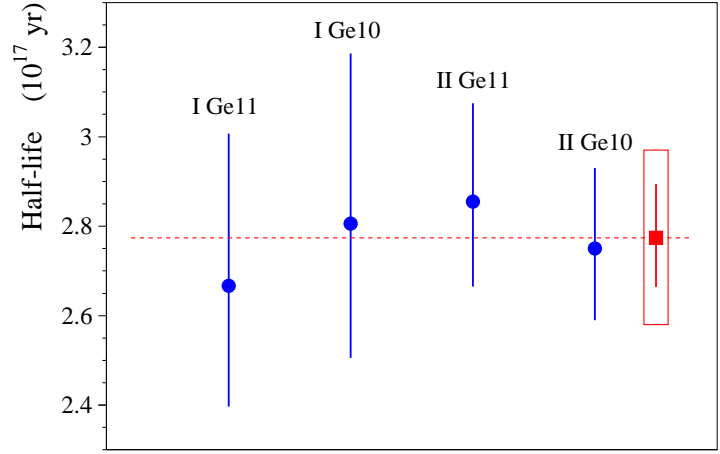,height=6.0cm}}
\caption{(Color online) Half-life of $^{50}$V relative to the
electron capture to the $2^+$ 1553.8 keV level of $^{50}$Ti
determined from the data of measurements with the detectors Ge10
and Ge11 in setups I and II (points, see also
Table~\ref{tab:T12-diff}). The final result of the present work,
obtained by analysis of the sum spectrum of the detectors in the
two setups, is shown by a square. The error bars represent the
statistical errors, while the box around the final value show the
errors calculated by summing in quadrature the statistical and
systematic uncertainties.}
 \label{fig:T12-diff}
\end{center}
\end{figure}

In addition to the $\approx0.2\%$ statistical uncertainty of the
Monte Carlo simulated detection efficiency we conservatively
assess a 4\%\footnote{See discussion of the difference between the
simulations and the experimental data used for the validation of
the simulations in Sec. \ref{sec:rad-cont}.} systematic
uncertainty on the calculated detection efficiency of the detector
system to the 1553.8 keV $\gamma$-ray quanta. An indirect
confirmation of a rather small systematic of the detection
efficiency can be seen in Table \ref{tab:T12-diff} and Fig.
\ref{fig:T12-diff} where the $T^{\mathrm{EC}}_{1/2}$ values
determined from the data of measurements with two different
detectors in setups I and II are presented. The difference between
the half-life values is well within the statistical errors, that
does demonstrate stability of the half-life result and its
independence neither on the detector nor the experimental setup.

Variation of the energy interval of fit from 1520--1540 keV
(starting point) to 1570--1585~keV (final point), changes
$T^{\mathrm{EC}}_{1/2}$ up to 1.1\%. Finally, we account 4.0\% for
uncertainty in the number of $^{50}$V nuclei in the sample due to
the accuracy of the representative isotopic abundance of the
isotope \cite{Meija:2016}. The summary of the systematic
uncertainties is given in Table~\ref{tab:syst}.

\begin{table}[htb]
\caption{Estimated systematic uncertainties of the EC decay
half-life (\%).}
\begin{center}
\begin{tabular}{|l|l|}
 \hline
 Number of $^{50}$V nuclei  & $4.0$ \\
 \hline
 Monte Carlo statistics     & $0.2$ \\
 \hline
 Monte Carlo systematic     & $4.0$ \\
 \hline
 Interval of fit            & $1.1$  \\
 \hline
 Total systematic uncertainty   & $5.8$ \\
 \hline
\end{tabular}
\end{center}
\label{tab:syst}
\end{table}

Adding all the systematic uncertainties in quadrature, the
half-life is

\begin{center}
 $T^{\mathrm{EC}}_{1/2}=[2.77^{+0.12}_{-0.11}(\mathrm{stat})\pm0.16(\mathrm{syst})]\times 10^{17}$ yr.
\end{center}

By summing in quadrature the statistical and systematic
uncertainties the half-life of $^{50}$Ti relative to the electron
capture to the $2^+$ 1553.8 keV excited level of $^{50}$Ti is

\begin{center}
$T^{\mathrm{EC}}_{1/2}=(2.77^{+0.20}_{-0.19})\times 10^{17}$ yr.
\end{center}

\begin{figure}[ht]
\begin{center}
 \mbox{\epsfig{figure=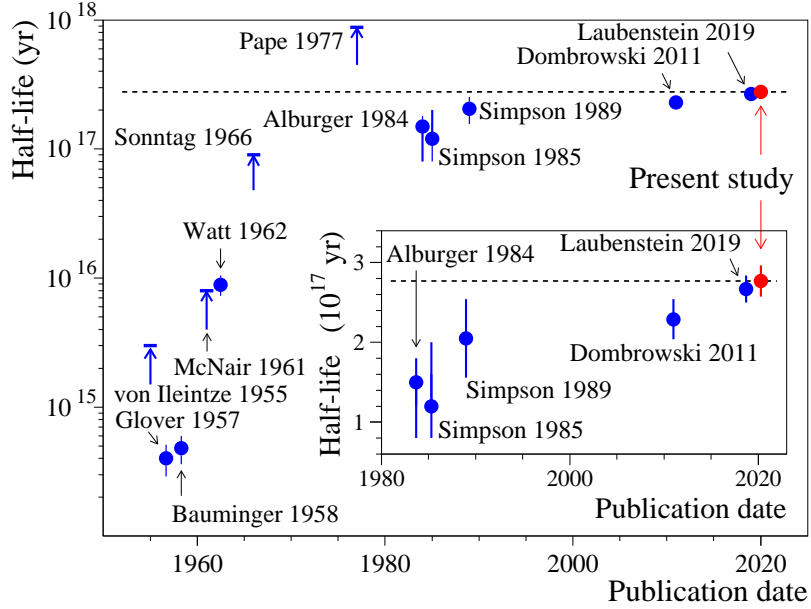,height=8.0cm}}
\caption{(Color online) A historical perspective of half-life of
$^{50}$V relative to the EC decay as a function of the publication
date (references to the publications are as follows: von Ileintze
1955: \cite{Ileintze:1955}, Glover 1957: \cite{Glover:1957},
Bauminger 1958: \cite{Bauminger:1958}, McNair 1961:
\cite{McNair:1961}, Watt 1962: \cite{Watt:1962}, Sonntag 1966:
\cite{Sonntag:1966}, Pape 1977: \cite{Pape:1977}, Alburger 1984:
\cite{Alburger:1984}, Simpson 1985: \cite{Simpson:1985}, Simpson
1989: \cite{Simpson:1989}, Dombrowski 2011:
\cite{Dombrowski:2011}, Laubenstein 2019:
\cite{Laubenstein:2019}). The results are presented by dots, while
the limits are shown by arrows. The early positive claims of EC
decay in $^{50}$V with too short half-lives were obtained with low
resolution detectors: proportional and NaI(Tl) scintillation
counters \cite{Glover:1957,Bauminger:1958}, NaI(Tl) scintillation
counter \cite{Watt:1962}. The half-lives measured with the help of
HPGe detectors in works
\cite{Alburger:1984,Simpson:1985,Simpson:1989,Dombrowski:2011,Laubenstein:2019}
and in the present study are in a reasonable agreement.}
 \label{fig:T12-hist}
\end{center}
\end{figure}

A historical perspective of half-life of $^{50}$V is presented in
Fig. \ref{fig:T12-hist}. It is interesting to note that early
experiments claimed too short half-lives. That can be explained,
first of all, by utilization of rather low energy resolution
detectors like proportional counters and NaI(Tl) scintillation
counters (see Table \ref{tab:hist}). Other possible reasons for
obtaining a too short half-life can be using nonpure samples, high
background with possible interferences of $\gamma$ rays of
different origin (including cosmogenic activation, since most of
the earlier experiments were performed in laboratories on the
ground level), less good electronics, stability problems of long
measurements. The latter point is especially crucial in conditions
of a poor energy resolution.

\clearpage

\subsection{Limit on $\beta^-$ decay of $^{50}$V to the $2^+$ 783.3 keV excited level of $^{50}$Cr}

There is no peak with energy $\approx~783$ keV in the sum energy
spectrum that can be interpreted as $\beta^-$ decay of $^{50}$V to
the $2^+$ 783.3 keV excited level of $^{50}$Cr. Thus, we have set
a lower half-life limit on the decay with the following formula:

\begin{equation}
\lim T_{1/2}=N~\ln 2 ~\sum{(\eta_i~ t_i)}/\lim S,
 \label{eq:limt12}
\end{equation}

\noindent where $N$ is the number of $^{50}$V nuclei in the
sample, $\eta_i$ and $t_i$ are detection efficiencies (for 783.3
keV $\gamma$-ray quanta) and times of measurement for the two
detectors in the two setups, and $\lim S$ is the number of events
of the effect searched for which can be excluded at a given
confidence level. The detection efficiencies for different
detectors were simulated with the help of the EGSnrc package
\cite{Kawrakow:2017,Lutter:2018}.

To estimate the value of $\lim S$ the sum energy spectrum with
exposure $(2.25\pm0.09)\times10^{22}$ nuclei of $^{50}$V$\times$yr
was fitted by a background model that includes the effect searched
for (a peak centered at 783.3 keV with a fixed FWHM~$=1.69$ keV),
several Gaussian peaks to describe background $\gamma$-ray peaks
of $^{138}$La, $^{212}$Bi (daughter of the $^{228}$Th subchain
from the $^{232}$Th chain), $^{214}$Bi and $^{214}$Pb (daughters
of $^{226}$Ra from the $^{238}$U chain), $^{228}$Ac (daughter of
the $^{228}$Ra subchain from the $^{232}$Th chain), $^{234m1}$Pa
(daughter of $^{238}$U), and a straight line to describe the
continuous background. While the areas and positions of intensive
peaks (766.4 keV of $^{234m1}$Pa, 768.4 keV of $^{214}$Bi, 785.4
keV of $^{212}$Bi, 788.7 keV of $^{138}$La, 795.0 keV of
$^{228}$Ac) were free parameters of the fit, the areas and
positions of weak peaks (772.3 keV and 782.1 keV of $^{228}$Ac,
786.0 keV of $^{214}$Pb, 786.3 keV of $^{234m1}$Pa, 786.4 keV of
$^{214}$Bi), superimposed on nearby intensive peaks, were fixed
taking into account their relative intensities in the sub-chains.
All the peak widths were fixed taking into account the dependence
of the energy resolution on energy of $\gamma$-ray quanta
(\ref{eq:fwhm}).

\begin{figure}[ht]
\begin{center}
 \mbox{\epsfig{figure=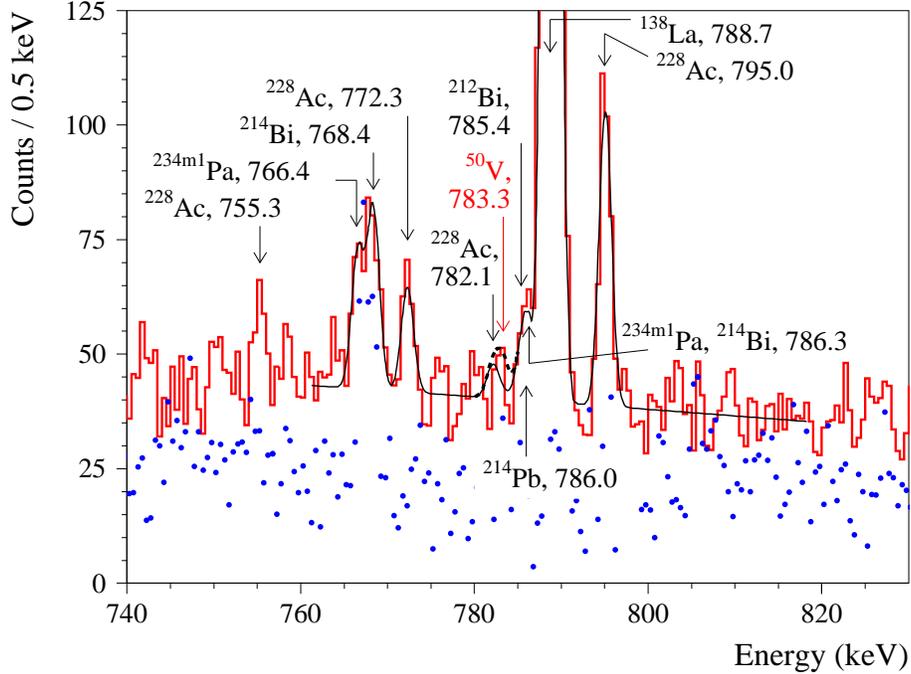,height=9.0cm}}
\caption{(Color online) Part of the sum energy spectrum
accumulated with the vanadium sample in the vicinity of the
expected $\beta^-$ decay 783.3 keV $\gamma$-ray peak. Fit of the
data by several $\gamma$-ray peaks and by a straight line to
describe the continuous background is shown by solid line, while
an excluded peak expected in the $\beta^-$ decay of $^{50}$V is
presented by dashed line. The background energy spectrum,
normalized on the time of measurements with the sample is shown by
dots. Energy of $\gamma$-ray peaks are in keV.}
 \label{fig:beta}
\end{center}
\end{figure}

The best fit, achieved in the energy interval 761--818 keV with
$\chi^2/$n.d.f.$=0.815$, returned an area $3.3\pm15.5$ counts in
an expected 783.3 keV peak that is no evidence of the effect
searched for.\footnote{The estimations of the $\lim S$ value
includes only the statistical uncertainty, and any systematic
contributions have not been considered.} The fit and excluded peak
are shown in Fig. \ref{fig:beta}. According to \cite{Feldman:1998}
we took $\lim S=28.7$ counts and, taking into account the
detection efficiencies to 783.3 keV $\gamma$-ray quanta (given in
Table \ref{tab:det-eff-783}), obtain the following limit on the
$\beta^-$ decay of $^{50}$V to the $2^+$ 783.3 keV excited level
of $^{50}$Cr:

\begin{center}
$T^{\mathrm{\beta}}_{1/2}\geq 8.9\times 10^{18}$ yr at 90\% C.L.
\end{center}

\begin{table}[ht]
\caption{Monte Carlo simulated full absorbtion peak detection
efficiencies for 783.3 keV $\gamma$-ray quanta for different
detectors in the two setups.}
\begin{center}
\begin{tabular}{|l|l|l|}

 \hline
 Setup  & Detector  & Detection efficiency \\
 \hline
 I      & Ge11      & 0.014986  \\
 I      & Ge10      & 0.018497 \\
 II     & Ge11      & 0.014140  \\
 II     & Ge10      & 0.018481  \\
 \hline
\end{tabular}
\label{tab:det-eff-783}
\end{center}
\end{table}

\clearpage

The limit is approximately two times weaker than the limit
$T_{1/2}^{\beta}\geq1.9\times10^{19}$ yr reported in
\cite{Laubenstein:2019}. The sensitivity of the present experiment
is lower mainly due to a rather high radioactive contamination of
the vanadium sample that produce background in the region of
interest.

Therefore, an advanced experiment should utilize a radio-pure
vanadium sample. A possibility of a deep purification of vanadium
from radioactive impurities has been demonstrated in
\cite{Laubenstein:2019}. Thus, aiming to estimate requirements to
experiments able to detect the decay, we assume a level of
background already achieved in setup II without sample (see Fig.
\ref{fig:set-up-II}). We consider two vanadium containing samples:
a metallic vanadium of the natural isotopic composition with the
sizes and geometry the same as in the present experiment, and a
second one in form of vanadium oxide (V$_2$O$_5$), enriched in the
isotope $^{50}$V to 50\%. We assume the bulk density of enriched
vanadium oxide sample to be 0.5 of the solid V$_2$O$_5$ density
(3.36 g/cm$^3$). To get the same number of $^{50}$V nuclei
($2.82\times10^{22}$), the size of the enriched sample was chosen
to be $\oslash50\times2.57$ mm, with a distance between the
detectors $H=3$ mm. Expected background counting rates and the
Monte Carlo simulated detection efficiencies of the Pacman setup
with the samples are given in Table \ref{tab:geom}.

\clearpage
\begin{table}[htbp]
\small \caption{Characteristics of experimental setups to estimate
sensitivity to the $\beta^-$ decay of $^{50}$V. $H$ denotes
distance between the detectors Ge10 and Ge11 (see Fig.
\ref{fig:set-ups}), $\mathrm{BG^{det}}$ is background counting
rate of the detectors (achieved in setup II without sample),
$\mathrm{BG^{EC}}$ is Monte Carlo simulated counting rate due to
the EC decay of $^{50}$V, $\eta_{783}$ is detection efficiency to
$\gamma$-ray quanta with energy 783.3 keV.}
\begin{center}
\begin{tabular}{|l|l|l|l|l|l|l|}
 \hline
 Sample,                                & \multicolumn{2}{c|}{$\mathrm{BG^{det}}$}  & \multicolumn{2}{c|}{$\mathrm{BG^{EC}}$}   & \multicolumn{2}{c|}{$\eta_{783}$} \\
 Experimental geometry                  & \multicolumn{2}{c|}{(counts/day/keV)}     & \multicolumn{2}{c|}{(counts/day/keV)}     & \multicolumn{2}{c|}{} \\
 \cline{2-7}
 ~                                      & Ge10              &  Ge11                 & Ge10                  &  Ge11             & Ge10      &  Ge11 \\
 \hline
 V metal, natural                       & ~                 & ~                     & ~                     & ~                 & ~         & ~  \\
 isotopic composition                   & ~                 & ~                     & ~                     & ~                 & ~         & ~  \\
 $\oslash100\times20$ mm, $H=21$ mm     & 0.1291(8)         & 0.1176(8)             & $0.0074$              & $0.0065$          & 0.01850   & 0.01499 \\
 \hline
 V$_2$O$_5$, enriched in $^{50}$V to 50\% & ~               & ~                     & ~                     & ~                 & ~         & ~  \\
 $\oslash50\times2.57$ mm, $H=3$ mm     & 0.1291(8)         & 0.1176(8)             & $0.0143$              & $0.0106$           & 0.05496 & 0.03716 \\

 \hline
\end{tabular}
%\normalsize
\end{center}
\label{tab:geom}
\end{table}

\begin{figure}[ht]
\begin{center}
 \mbox{\epsfig{figure=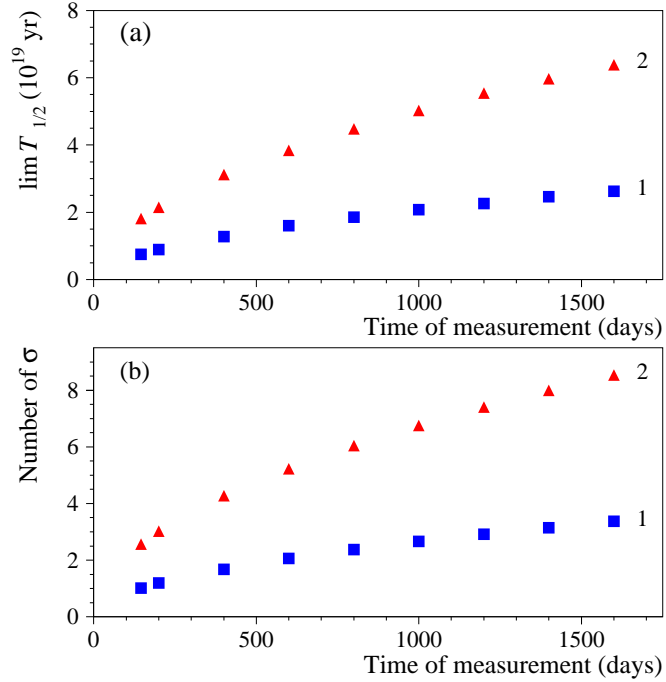,height=9.0cm}}
\caption{(Color online) Sensitivity of possible experiments to
detect the $\beta^-$ decay of $^{50}$V [expressed as: (a) a lower
half-life limit at 90\% C.L.; (b) a number of $\sigma$ for
accuracy of the expected 783.3 keV peak area, assuming the
half-life $T_{1/2}^{\beta^-}=2\times10^{19}$ yr] depending on time
of measurement in two experimental conditions: (1) in the geometry
of the present experiment (with a V-sample $\oslash100\times20$ mm
and distance between the detectors Ge10 and Ge11 $H=21$ mm); (2)
with a V$_2$O$_5$-sample enriched in the isotope $^{50}$V to 50\%
with sizes $\oslash50\times2.57$ mm and $H=3$ mm. Only background
without sample together with contribution due to the EC decay of
$^{50}$V are assumed.}
 \label{fig:beta-sens}
\end{center}
\end{figure}

The background of the detectors dominates in the experimental
conditions, with the contribution from the EC process in $^{50}$V
an order of magnitude smaller. While the assumed enriched source
contains the same number of $^{50}$V nuclei as the metallic one
with the natural isotopic composition, the detection efficiency
with the enriched source is about three times higher. As a result,
an experiment with enriched source has a higher sensitivity [see
Fig. \ref{fig:beta-sens}, (a)]. Moreover, utilization of enriched
$^{50}$V would allow to observe clearly the $\beta^-$ decay of
$^{50}$V (assuming the theoretically predicted half-life
$T_{1/2}^{\beta^-}=2\times10^{19}$ yr \cite{Haaranen:2014}) with a
$3\sigma$ accuracy over about 200 d of data taking, while an
experiment utilizing a V-sample of natural isotopic composition
needs more than three years to detect the process with a similar
accuracy (see Fig. \ref{fig:beta-sens}, (b)).

\clearpage

\section{CONCLUSIONS}

The half-life of $^{50}$V relative to the EC to the $2^+$ 1553.8
keV level of $^{50}$Ti is measured as
$T^{\mathrm{EC}}_{1/2}=(2.77^{+0.20}_{-0.19})\times 10^{17}$~yr.
The value is in agreement with the result of the recent experiment
\cite{Laubenstein:2019} and the theoretical predictions
\cite{Haaranen:2014}. The $\beta^-$ decay of $^{50}$V to the $2^+$
783.3 keV level of $^{50}$Cr is limited as
$T^{\beta}_{1/2}\geq8.9\times 10^{18}$ yr at 90\% C.L. The limit
is about 2 times weaker than that set in the work
\cite{Laubenstein:2019}. Further improvement of the experiment
sensitivity could be achieved by utilization of highly purified
vanadium samples. Moreover, using of a sample enriched in $^{50}$V
would allow detection of the $\beta^-$ decay. The accuracy of
$T^{\mathrm{EC}}_{1/2}$ will also be improved with a source
enriched in $^{50}$V both thanks to improvement of statistics and
reduction of the uncertainty in the $^{50}$V isotopic abundance.

\section{ACKNOWLEDGMENTS}

This work received support from the EC-JRC open access scheme
EUFRAT under Horizon-2020, project No. 22-14. D.V.K. and O.G.P.
were supported in part by the project ``Investigation of double
beta decay, rare alpha and beta decays'' of the program of the
National Academy of Sciences of Ukraine ``Laboratory of young
scientists'' (the grant number 0120U101838). F.A.D. greatly
acknowledges the Government of Ukraine for the quarantine measures
that have been taken against the Coronavirus disease 2019 that
substantially reduced much unnecessary bureaucratic work.


\begin{thebibliography}{99}

 \bibitem{Meija:2016} J.~Meija {\it et al}., Isotopic compositions of the elements 2013 (IUPAC Technical Report), Pure Appl. Chem. 88 (2016) 293.

 \bibitem{Wang:2017} M.~Wang {\it et al}., The AME2016 atomic mass evaluation, Chin. Phys. C 41 (2017) 030003.

 \bibitem{Ileintze:1955} J. von Ileintze, Zur Frage der nat\"urlichen Radioaktivit\"at des V$^{50}$, In$^{113}$ und Te$^{123}$,  Z. Naturforschung 10a (1955) 77.

 \bibitem{Glover:1957} R. N.~Glover and D. E.~Watt, A search for natural radioactivity in vanadium, Philos. Mag. 2 (1957) 697.

 \bibitem{Bauminger:1958} E. R.~Bauminger and S. G.~Cohen, Natural radioactivity of V$^{50}$ and Ta$^{180}$, Phys. Rev. 110 (1958) 953.

 \bibitem{McNair:1961} A.~McNair, The half-life of vanadium-50, Philos. Mag. 6 (1961) 559.

 \bibitem{Watt:1962} D. E.~Watt and R. L. G.~Keith, The half-life of $^{50}$V, Nucl. Phys. 29 (1962) 648.

 \bibitem{Sonntag:1966} Ch.~Sonntag and K. O.~M\"unnich, A search for the decay of vanadium-50 with a low-level gamma-spectrometer, Z. Physik 197 (1966) 300.

 \bibitem{Pape:1977} A.~Pape, S. M.~Refaei, and J. C.~Sens, $^{50}$V half-life limit, Phys. Rev. C 15 (1977) 1937.

 \bibitem{Alburger:1984} D. E.~Alburger, E. K.~Warburton, and J. B.~Cumming, Decay of $^{50}$V, Phys. Rev. C 29 (1984) 2294.

 \bibitem{Simpson:1985} J. J.~Simpson, P.~Jagam, and A. A.~Pilt, Electron capture decay rate of $^{50}$V, Phys. Rev. C 31 (1985) 575.

 \bibitem{Simpson:1989} J. J.~Simpson, P.~Moorhouse, and P.~Jagam, Decay of $^{50}$V, Phys. Rev. C 39 (1989 ) 2367.

 \bibitem{Dombrowski:2011} H.~Dombrowski, S.~Neumaier, and K.~Zuber, Precision half-life measurement of the 4-fold forbidden $\beta$ decay of $^{50}$V, Phys. Rev. C 83 (2011) 054322.

 \bibitem{Laubenstein:2019} M.~Laubenstein, B.~Lehnert, S.~S.~Nagorny, S.~Nisi, and K.~Zuber, New investigation of half-lives for the decay modes of $^{50}$V, Phys. Rev. C 99 (2019) 045501.

 \bibitem{Belli:2019} P.~Belli {\it et al}., Experimental searches for rare alpha and beta decays, Eur. Phys. J. A 55 (2019) 140.

 \bibitem{Haaranen:2014} M.~Haaranen, P. C.~Srivastava, J.~Suhonen, and K.~Zuber, $\beta$-decay half-life of $^{50}$V calculated by the shell
 model, Phys. Rev. C 90 (2014) 044314.

 \bibitem{Barea:2014} J. Barea, J. Kotila, and F. Iachello, Nuclear matrix elements for double-$\beta$ decay, Phys. Rev. C 87 (2013) 014315.

 \bibitem{DellOro:2014} S. Dell'Oro, S. Marcocci, and F. Vissani, New expectations and uncertainties on neutrinoless double beta decay, Phys. Rev. D 90 (2014) 033005.

 \bibitem{Suhonen:2017a} J. T.~Suhonen, Impact of the quenching of $g_A$ on the sensitivity of $0\nu2\beta$ experiments, Phys. Rev. C 96 (2017) 055501.

 \bibitem{Suhonen:2017b} J. T. Suhonen, Value of the axial-vector coupling strength in $\beta$ and $\beta\beta$ decays: A review, Front. in Phys. 5 (2017) 55.

 \bibitem{Ejiri:2019} H.~Ejiri, J.~Suhonen, and K.~Zuber, Neutrino-nuclear responses for astro-neutrinos, single beta decays and double beta decays, Phys. Rep. 797 (2019) 1.

 \bibitem{Lutter:2013} G.~Lutter {\it et al}., A new versatile underground gamma-ray spectrometry
 system, Appl. Radiat. Isot. 81 (2013) 81.

 \bibitem{Hult:2013} M.~Hult {\it et al}., Comparison of background in underground HPGe-detectors in different lead shield
 configurations, Appl. Radiat. Isot. 81 (2013) 103.

 \bibitem{Kawrakow:2017} I.~Kawrakow {\it et al}., The EGSnrc Code System: Monte Carlo Simulation of Electron and Photon Transport. Technical Report No. PIRS-701, National Research Council Canada (2017).

 \bibitem{Lutter:2018} G.~Lutter, M.~Hult, G.~Marissens, H.~Stroh, and F.~Tzika, A gamma-ray spectrometry analysis software environment, Appl. Rad. Isot. 134 (2018) 200.

 \bibitem{Chen:2019} J. Chen and B. Singh, Nuclear Data Sheets for A=50, Nucl. Data Sheets 157 (2019) 1.

 \bibitem{Feldman:1998} G. J.~Feldman and R. D.~Cousins, Unified approach to the classical statistical analysis of small signals, Phys. Rev. D 57 (1998) 3873.

\end{thebibliography}
\end{document}